\documentclass[11pt,a4paper]{article}
\usepackage{enumerate}
\usepackage{fancyhdr}
\usepackage{url}
\usepackage{mathrsfs}
\usepackage{amsmath}
\usepackage{amsfonts}
\usepackage{amssymb}
\usepackage{amsthm}
\usepackage{graphicx}
\usepackage{pst-all}
\usepackage{comment}
\usepackage[toc,page]{appendix}
\usepackage{multicol}
\newtheorem{defn}{Definition}

\newtheorem*{so1}{SO1}
\newtheorem*{so2}{SO2}
\newtheorem*{finiteso1}{Finite SO1}
\newtheorem*{finiteso2}{Finite SO2}

\newtheorem{Proposition}{Proposition}

\newtheorem{Remark}{Remark}
\begin{document}
\hyphenation{com-mon Rei-chen-bach Rei-chen-ba-chian}


\title{Distinguishing causality principles}
\author{Mikl{\'o}s R{\'e}dei\\ {\normalsize{Department of Philosophy, Logic and Scientific Method}}\\ {\normalsize{London School of Economics and Political Science}}\\ {\normalsize{Houghton Street, London WC2A 2AE, England}}\\ {\normalsize{\tt M.Redei@lse.ac.uk}}\\ \\ I{\~n}aki San Pedro\\ {\normalsize{Department of Logic and Philosophy of Science}}\\ {\normalsize{Complutense University Madrid}}\\ {\normalsize{Ciudad Universitaria, 28040 Madrid, Spain}}\\ {\normalsize{\tt inaki.sanpedro@filos.ucm.es}}}
\date{}
\maketitle

\renewcommand{\abstractname}{Abstract}
\begin{abstract}
We distinguish two sub-types of each of the two causality principles formulated in connection with the Common Cause Principle in \cite{Henson2005} and raise and investigate the problem of logical relations among the resulting four causality principles. Based in part on the analysis of the status of these four principles in algebraic quantum field theory we will argue that the four causal principles are non-equivalent.
\end{abstract}
\vspace{-.6cm}


\pagestyle{fancy}
\renewcommand{\sectionmark}[1]{\markboth{#1}{}}
\fancyhf{}
\fancyfoot[C,C]{\thepage}
\fancyhead[C]{\it \leftmark}
\renewcommand{\headrulewidth}{0pt}
\renewcommand{\footrulewidth}{0pt}
\addtolength{\headheight}{2.5pt}
\fancypagestyle{plain}{
\fancyhead{}
\renewcommand{\headrulewidth}{0pt}
}

\pagestyle{fancy}
\section{Introduction}
Distinguishing and comparing carefully causality principles is important when it comes to the question of whether a fundamental physical theory is compatible with ``causality''. Motivated to a large extent by the well-known difficulties that stand in the way of explaining quantum (EPR) correlations causally, Henson \cite{Henson2005} proposed two apparently different causality principles (SO1 and SO2 in the notation of \cite{Henson2005}) and proved that the two principles are in fact equivalent. The equivalence claim states, roughly, that two requirements about the localizability of a common cause of a correlation between causally disjoint events $A$ and $B$ are equivalent: localizability within the {\em mutual\/} causal past of $A$ and $B$ (this is SO1) and localizability within the {\em joint\/} causal past of $A$ and $B$ (this is SO2). The aim of this paper is to distinguish further two sub-types of both SO1 and SO2 and to raise and investigate the problem of logical relations among the resulting four causality principles.

The need to distinguish the further types arises from the fact that the SO1 $\Leftrightarrow$ SO2 equivalence seems counterintuitive: the mutual past being part of the joint past, common causes localized in the mutual past are localized in the joint past as well, hence SO1 $\Rightarrow$ SO2 corresponds to our intuition; however, the converse implication SO1 $\Leftarrow$ SO2 does not. The intuition about this asymmetry is supported by specific results about existence of localized common causes in relativistic quantum field theory: The equivalence SO1 $\Leftrightarrow$ SO2 indicates that algebraic relativistic quantum field theory, which predicts correlations between observables localized in causally disjoint regions, would satisfy the Common Cause Principle in the sense of providing causal explanations of spacelike correlations in terms of common causes localized in the {\em mutual\/} causal past of the spacelike separated spacetime regions containing the correlated observables because common causes localized in the {\em joint\/} pasts have been shown to exist \cite{Redei-Summers2002}, \cite{Redei-Summers2005corr}. Given that no proof of stricter localizability of common causes in algebraic quantum field theory could be given so far, and in view of the fact that recent further investigations of the status of the common cause principle in lattice quantum field theory seem to indicate that the two localizability conditions are {\em not\/} equivalent \cite{Hofer-Vecsernyes2012}, a careful scrutiny of the equivalence claim and of its proof is called for.

A look at the proof of the equivalence claim reveals that the crucial feature of SO2 that the proof of the implication SO2 $\Rightarrow$ SO1 is based on is that SO2 allows to consider correlations between events associated with ``causally infinite'' regions and requires such correlations to have common causes localized in the joint past of the causally infinite  regions. We will argue that this makes SO2 extremely strong and that it is both physically and intuitively justified to distinguish such ``infinite SO2'' from a weaker, ``finite SO2'' principle that restricts the correlations for which common causes are required to exist to  correlations between events in ``causally finite'' regions (see Definition \ref{def:causalfiniteness} for causally infinite and causally finite regions). A similar distinction can be made in connection with SO1, and thus one has four distinct causality principles: finite and infinite SO2, on one hand, and finite and infinite SO1 on the other. The question arises then what the relation of these four principles are. This seems to be a difficult problem that remains largely open. Henson proves the equivalence of {\emph{infinite} SO2 and \emph{infinite} SO1 ---although we detect what appears a slight gap in the proof of the implication [\emph{infinite} SO1 $\Rightarrow$ \emph{infinite} SO2], see the end of Section~\ref{sec:equivalent?}. This gap questions whether the implication [{\em finite} SO1 $\Rightarrow$ {\em finite} SO2] holds, but we do not have a counterexample to the implication. It will be shown however that Henson's method does \emph{not} prove the implication [{\em finite} SO2 $\Rightarrow$ {\em finite} SO1], and that it does {\em not} prove the implication [\textit{finite} SO2 $\Rightarrow$ \textit{infinite} SO1] either, and we will argue, although we are unable to provide a strict formal proof, that the {\emph{finite} SO2 does \emph{not} entail even the \emph{finite} SO1.

Henson's formulates the causality principles SO1 and SO2 within a very elaborate formal framework that links explicitly probabilistic concepts to spatiotemporal notions.  Section~\ref{sec:hensons_framework} recalls briefly this framework and some definitions needed to formulate the two causality principles SO1 and SO2, which will be stated explicitly in Section~\ref{sec:CCP} together with Henson's proof of their equivalence. Section~\ref{sec:finite-infinite} analyzes the equivalence proof and motivates the distinction between finite and infinite versions of SO1 and SO2. Finally, Section~\ref{sec:equivalent?} discusses the problem of relation of the principles, arguing that they are not equivalent.

\section{Events, probabilities and least domain of decidability
\label{sec:hensons_framework}}
Henson creates a formal framework in order to relate probabilistic concepts to spatiotemporal notions, the latter understood as including concepts of non-probabilistic causality. Here we recall this framework following the notations of \cite{Henson2005} closely.

Let $\mathbb{S}$ be some spacetime structure. Henson leaves the nature of $\mathbb{S}$ unspecified except that he assumes a causal (partial) order $<$ defined on $\mathbb{S}$ that expresses causal precedence between elements $x,y$ in $\mathbb{S}$: if $x<y$ then $x$ is in the causal past of $y$.\footnote{Henson lists as examples of such causally ordered sets weakly causal Lorentzian manifolds and causal sets~\cite[p.\ 523]{Henson2005}. Weakly causal Lorentzian manifolds are manifolds that contain no causally closed curves; an example is the Minkowski spacetime. If the causal partial order $<$ defined in $\mathbb{S}$ is transitive, acyclic and locally finite, then $\mathbb{S}$ is called a \emph{causal set}. Causal sets so defined then constitute the discreet equivalent to spacetime manifolds equipped with causal structure. (Discreetness is introduced by the locally finite property of the order relation in a causal set, i.e.\ $\forall x, z \in \mathbb{S},\ \mathrm{card}\{ y \in \mathbb{S}:\ x < y < z \} < \infty$. See~\cite{Sorkin2003} for details.)} Given $(\mathbb{S},<)$ the past $J^-(x)$ of $x \in \mathscr{X}$, is defined as the set of all points that precedes $x$ in the causal order, plus $x$ itself, i.e.\
\[
J^- (x) = \{ x_p \in \mathbb{S}: x_p < x \} \cup \{x\}
\]

The past of $\mathscr{X}\subset\mathbb{S}$ is then defined \cite[p.\ 523]{Henson2005} as the union of the pasts of all points in the region:
\[
J^- (\mathscr{X}) = \bigcup_{x \in \mathscr{X}} J^- (x)
\]

Given two regions $\mathscr{X}$ and $\mathscr{Y}$ in $(\mathbb{S},<)$ one can define the following regions:
\begin{description}
\item[Mutual past:] $\mathscr{P}_1(\mathscr{X},\mathscr{Y})=J^-(\mathscr{X}) \cap J^-(\mathscr{Y})$
\item[Joint past:]  $J^-(\mathscr{X}) \cup J^-(\mathscr{Y})$
\item[Truncated joint past:] $\mathscr{P}_2(\mathscr{X},\mathscr{Y})=(J^-(\mathscr{X})\cup J^-(\mathscr{Y}))\setminus (\mathscr{X}\cup\mathscr{Y})$
\end{description}
Note that the terminology in \cite{Henson2005} is slightly different: Henson calls the \emph{truncated} joint past $\mathscr{P}_2(\mathscr{X},\mathscr{Y})$ ``joint past'' and does not define joint past explicitly. We will use the terminology introduced above.

Two regions $\mathscr{X}$ and $\mathscr{Y}$ are defined to be space-like separated\label{def:space-like_sep} if
\begin{align*}
J^- (\mathscr{X}) \cap \mathscr{Y} & = \emptyset, \\
\mathscr{X} \cap  J^- (\mathscr{Y}) & = \emptyset.
\end{align*}

In the notations of \cite{Henson2005}, a classical probability measure space is a triplet $(\Omega, \Sigma,\mu)$ where $\Omega$ is ``the space of all possible histories of the system'' \cite [p.\ 521]{Henson2005} to be described probabilistically, $\Sigma$ is an ``appropriate $\sigma$ algebra of subsets of $\Omega$'' \cite[p.\ 521]{Henson2005} and $\mu\colon \Sigma\to\mathbb{R}$ is a probability measure ``obeying the usual axioms'' \cite[p.\ 521]{Henson2005}. The link between $(\Omega, \Sigma,\mu)$ and spatiotemporal concepts is given by a map
\[
dom: \Sigma \rightarrow \mathrm{pow}(\mathbb{S})
\]
where $\mathrm{pow}(\mathbb{S})$ denotes the power set of $\mathbb{S}$. The region  $dom(A)$ is called the {\em least domain of decidability} of event $A\in\Sigma$ -- intuitively, $dom(A)$ is the unique spacetime region such that if we know all the properties of the history in it we can decide, without any further information, whether $A$ has occurred or not.

The properties of $dom$ are fixed axiomatically by formulating requirements that it should satisfy. Following \cite[p.\ 524]{Henson2005} we list the axioms below. \\

\noindent\textbf{Properties of $dom$:}\\
For all countable subsets $\Lambda\subset\Sigma$ we have
\begin{enumerate}
\item If $dom(X) \cap dom(Y) = \emptyset \quad \forall X, Y \in \Lambda$, such that $X \neq Y$ then
\[
dom(\bigcap_{X \in \Lambda} X) = \bigsqcup_{X \in \Lambda} dom(X)
\]
\noindent where $\bigsqcup$ stands for the `disjoint union'.

\item If $dom(X) = dom(Y) \quad \forall X, Y \in \Lambda$ then
\[
dom(\bigcap_{X \in \Lambda} X) \subset dom(Y), \quad \forall Y \in \Lambda.
\]

\item $dom(X^c) = dom(X), \quad \forall X \in \Sigma$.

\item $\forall Z \in \Sigma$ such that $dom(Z) = \mathscr{X} \sqcup \mathscr{Y}$, $Z$ is a member of the $\sigma$-algebra generated by $\Gamma(\mathscr{X}) \cup \Gamma(\mathscr{Y})$, where $\Gamma(\mathscr{X}) = \{ X \in \Sigma : dom(X) \subset \mathscr{X} \}$.
\end{enumerate}
A further notion that will be important in the discussion of the common cause principle is the {\em full specification of a region}:
\begin{defn}\label{deffullspecification}
Event $F\in\Sigma$ is a full specification of a region $\mathscr{R}$ if $dom(F)\subset\mathscr{R}$, and for all $X\in\Sigma$ it holds that, if $dom(X)\subset\mathscr{R}$, then either $F\subset X$ or $F\subset X^c$.
\end{defn}
$\Phi(\mathscr{R})$ denotes the set of full specifications of region $\mathscr{R}$. Henson proves the following proposition:
\begin{Proposition}[Lemma 3 in \cite{Henson2005}]\label{Prop:partition}
The full specification of any region is a partition of $\Omega$.
\end{Proposition}

\section{Two Common Cause Principles: SO1 and SO2 and the Equivalence Claim}\label{sec:CCP}
The next definition of common cause was given by Reichenbach \cite{Reichenbach1956} and is standard in the literature.
\begin{defn}\label{def:commoncause}
Assume that $A,B\in\Sigma$ are positively correlated in $\mu$:
\begin{align}
\label{def:corr}
\mu(A \cap B) & > \mu (A) \mu (B)
\end{align}
Event $C\in\Sigma$ is a (Reichenbachian) common cause of the correlation \eqref{def:corr} if it satisfies the following conditions
\begin{align}
\mu(A \cap B \vert C) & = \mu (A \vert C) \mu (B \vert C) \\
\mu(A \cap B \vert C^c) & = \mu (A \vert C^c) \mu (B \vert C^c) \\
\mu(A \cap C) & > \mu (A \cap C^c)\label{relevance1}\\
\mu(B \cap C) & > \mu (B \cap C^c)\label{relevance2}
\end{align}
\end{defn}
Reichenbach also formulated what is called the

\smallskip\noindent
{\bf Common Cause Principle}: If two events $A,B$ are correlated then either there is a direct causal link between $A$ and $B$ that is responsible for the correlation, or, if $A$ and $B$ are causally independent, then there exists a common cause that explains the correlation.

\smallskip

Viewed from the perspective of the Common Cause Principle, the definition of common cause (Definition \ref{def:commoncause}) is incomplete for two reasons:
\begin{enumerate}
\item The definition does not contain an explicit requirement about the causal {\em independence} of $A$ and $B$.
\item It is part of the intuition about the Common Cause Principle that the common cause $C$ should be an event that can causally affect events $A$ and $B$; however, there is no explicit expression of this type of non-probabilistic causal link between $C$ and $A$ and $C$ and $B$ in the definition of common cause.
\end{enumerate}
How should one transform the two informal conditions (i) and (ii) above into precise and technically explicit conditions in order to obtain a formulation of the Common Cause Principle that can be formally analyzed? In our view this is the deepest and most important problem in connection with the Common Cause Principle, which lies at the heart of most of the debates about the Common Cause Principle. It is here that Henson's framework and the notion of least domain of decidability offers a solution. Before recalling how, we need to make two remarks about possible modifications of the notion of common cause:

The first remark is that there are arguments in favor of {\em weakening} the notion of common cause in the following sense: It is conceivable that a correlation is due not to a single common cause but a whole set of ``partial common causes''. This idea leads to the notion of common cause {\em system}, which was analyzed in a number of publications (see \cite{Uffink1999}, \cite{Hofer-Redei2004},  \cite{Hofer-Redei2006}, \cite{Wronski-Marczyk2010}). A common cause system is a partition $\{C_i\ :\ i\in I\}$ of $\Omega$ such that
\begin{equation}\label{arnyekol}
\mu(A\cap B|C_i)=\mu(A|C_i)p(B|C_i) \qquad \mbox{for all \ }i\in I\\
\end{equation}
\begin{equation}\label{pozitiv}
[\mu(A\vert C_i )-\mu(A\vert C_j)] [\mu(B\vert C_i)-\mu(B\vert C_j)]>0 \ \ (i\not=j)
\end{equation}
Equation \eqref{arnyekol} is called the {\em screening-off} condition, \eqref{pozitiv} is the analogue of the statistical relevance conditions \eqref{relevance1}-\eqref{relevance2} (see \cite{Hofer-Redei2004},  \cite{Hofer-Redei2006} for a further motivation of condition \eqref{pozitiv}).

The second remark is that when it comes to proving that common causes cannot exist for certain correlations, one would like to prove such ``no-go'' theorems under the weakest conditions; hence, frequently the general statistical relevance condition \eqref{pozitiv} is dropped from the definition of common cause system ---this is what Henson also does in the formulation of the two causality principles  SO1 and SO2. These principles are as follows.
\begin{so1}
For all events $A$ and $B$ with $dom(A) \subset \mathscr{A}$ and $dom(B) \subset \mathscr{B}$, if $\mathscr{A}$ and $ \mathscr{B}$ are space-like separated regions, then
\begin{align}
& \mu(A \cap B \vert C) = \mu(A \vert C) \mu(B \vert C) \quad \forall C \in \Phi(\mathscr{P}_1(\mathscr{A},\mathscr{B})).
\label{eq:so1}
\end{align}
\end{so1}
\begin{so2}
For all events $A$ and $B$ with $dom(A) \subset \mathscr{A}$ and $dom(B) \subset \mathscr{B}$, if $\mathscr{A}$ and $\mathscr{B}$ are space-like separated regions then:
\begin{align}
& \mu(A \cap B \vert C) = \mu(A \vert C) \mu(B \vert C) \quad \forall C \in \Phi(\mathscr{P}_2(\mathscr{A},\mathscr{B})).
\label{eq:so2}
\end{align}
\end{so2}
Some remarks on SO1 and SO2 are in order:
\begin{Remark}
\noindent
{\rm
\begin{enumerate}
\item Note that the $C$'s referred to in SO1 and SO2 do form a partition because they are full specifications of regions and these form a partition by Proposition \ref{Prop:partition}; hence both SO1 and SO2 state that correlations between events with causally disjoint domains of decidability have a common cause system.
\item The fact that $C$'s are required to be {\em full specifications} in the respective regions entails (and thus is motivated by) the fact that this way ``Simpson events'', events that could re-introduce the correlation between $A$ and $B$ after they have been screened off by $C$'s, are excluded (see \cite[p.\ 526-527]{Henson2005} for a further discussion of the ``Simpson problem'').
\item The two principles differ in which past they require the postulated screening-off partition to be located: The first principle requires the common cause system to lie in the \emph{mutual past} of $A$ and $B$: recall that $\mathscr{P}_1(\mathscr{A},\mathscr{B}) = J^-(\mathscr{A}) \cap J^-(\mathscr{B})$. The second principle requires the common cause system to be localized in the \emph{truncated joint past} of $A$ and $B$: $\mathscr{P}_2(\mathscr{A},\mathscr{B}) = (J^-(\mathscr{A}) \cup J^-(\mathscr{B})) \setminus (\mathscr{A} \cup \mathscr{B})$.
\item There is a deceptive dissimilarity in requiring the common cause system in SO1 to be in the {\em untruncated} mutual past and to be in the \emph{truncated} joint past in SO2; however, since $\mathscr{A}$ and $ \mathscr{B}$ are assumed to space-like separated the untruncated and truncated mutual pasts of $\mathscr{A}$ and $ \mathscr{B}$ coincide; hence there is no need to require explicitly the common cause system to be in the \emph{truncated} mutual past. But one has to do so explicitly in SO2 because, intuitively, we do \emph{not} want the common cause system to be in the regions where the correlated events are. Requiring the common cause system to be in the \emph{truncated} mutual past excludes for instance the common cause system to be the partition generated by $A$ and $B$.
\item The difference between the two localizability regions in SO1 and SO2 is that
\begin{itemize}
\item The mutual past is the set of points in $\mathbb{S}$ {\em each\/} of which is causally related to {\em at least one\/} point in {\em both\/}  $\mathscr{A}$ and $\mathscr{B}$.
\item The truncated joint past consist of the set of points in $\mathbb{S}$ \emph{none} of which is in either $\mathscr{A}$ or in  $\mathscr{B}$, and  \emph{each} of which is causally related to {\em at least one\/} point in {\em either\/} $\mathscr{A}$ {\em or\/} $\mathscr{B}$.
\end{itemize}
\end{enumerate}
}
\end{Remark}
The description (v) above of the mutual and joint pasts indicates that SO1 and SO2 are different: SO2 seems a weaker principle than SO1.
Henson claims however that SO1 and SO2 are equivalent (Corollary 2 in \cite{Henson2005}).
We now sketch the proof of this equivalence claim.
In the proof the following two regions $\mathscr{X}$ and $\mathscr{Y}$ will be referred to:
\begin{eqnarray}
\mathscr{X}& = &\left[ J^-(\mathscr{A}) \backslash \mathscr{A} \right] \backslash J^-(\mathscr{B})
\label{def:region_x}\\
\mathscr{Y} &=& \left[ J^-(\mathscr{B}) \backslash \mathscr{B} \right] \backslash J^-(\mathscr{A})
\label{def:region_y}
\end{eqnarray}
\begin{description}
\item[(I).]Proof of \textbf{SO1 $\Rightarrow$ SO2}\label{proof:so1so2}\\
Assume SO1. Then for $A$ and $B$ such that $dom(A) \subset \mathscr{A}$, $dom(B) \subset \mathscr{B}$ and $\mathscr{A}$ and $\mathscr{B}$ space-like separated we have:
\begin{equation}\label{eqinproof1}
\mu(A \cap B \vert C) = \mu(A \vert C) \mu(B \vert C) \quad \forall C \in \Phi(\mathscr{P}_1(\mathscr{A},\mathscr{B})),
\end{equation}
Let $X$ and $Y$ be \emph{full specifications} of regions $\mathscr{X}$ and $\mathscr{Y}$, respectively, and consider the following four pairs of events
\begin{equation}\label{pairs}
\{ (A \cap X), (B \cap Y) \} \quad  \{ (A \cap X), B \} \quad  \{ A, (B \cap Y) \} \quad  \{ X, Y \}
\end{equation}

It is easy to show that the two events in each of the above pairs occur in regions $\{ (\mathscr{A} \cup \mathscr{X}), (\mathscr{B} \cup \mathscr{Y})\}$, which are space-like separated; on the other hand the \emph{mutual past} of
$(\mathscr{A} \cup \mathscr{X})$ and $(\mathscr{B} \cup \mathscr{Y})$ is equal to the mutual past of $\mathscr{A}$ and $\mathscr{B}$:
\[
J^-(\mathscr{A} \cup \mathscr{X}) \cap J^-(\mathscr{B} \cup \mathscr{Y}) =  J^-(\mathscr{A}) \cap J^-(\mathscr{B})
\]

Thus SO1 can be applied to each pair in eq.\ \eqref{pairs} and we obtain that any full specification $C$ of $\Phi(\mathscr{P}_1(\mathscr{A},\mathscr{B}))$ screens off the correlation between the elements in every of these pairs.

A probability Lemma~\cite[Lemma~4, p.\ 528]{Henson2005} is then invoked to show that $\forall X \in \Phi(\mathscr{X}), \ \forall Y \in \Phi(\mathscr{Y})$ and $C \in \Phi(\mathscr{P}_1(\mathscr{A},\mathscr{B}))$ the following relation follows:
\[
 \mu (A \vert X \cap Y \cap C) \mu (B \vert X \cap Y \cap C) = \mu (A \cap B \vert X \cap Y \cap C).
\]

On the other hand, a full specification $F$ of a region $\mathscr{R}$ defined as the disjoint union of some finite set of regions $\mathscr{A}_i$, i.e.\ $\mathscr{R} = \sqcup_i \mathscr{A}_i$, can be written as $F = \cap_i A_i$, where $A_i$ is a full specification of $\mathscr{A}_i$~\cite[Corollary~1, p.\ 525]{Henson2005}.

Then, since $X$, $Y$ and $C$ are full specifications of $\mathscr{X}$, $\mathscr{Y}$ and $\Phi(\mathscr{P}_1(\mathscr{A},\mathscr{B}))$ respectively, i.e.\ $X \in \Phi(\mathscr{X})$, $Y \in \Phi(\mathscr{Y})$ and $C \in \Phi(\mathscr{P}_1(\mathscr{A},\mathscr{B}))$, and since $\mathscr{P}_2(\mathscr{A},\mathscr{B})= \mathscr{X} \sqcup \mathscr{Y} \sqcup \mathscr{P}_1$, we have that
\[
C \cap X \cap Y \in \Phi(\mathscr{P}_2(\mathscr{A},\mathscr{B})).
\]

It is concluded then that SO1 implies SO2.
\item[(II).]\textbf{SO2 $\Rightarrow$ SO1}\\
Assume now SO2. Thus if $A$ and $B$ are such that $dom(A) \subset \mathscr{A}$, $dom(B) \subset \mathscr{B}$ and $\mathscr{A}$ and $\mathscr{B}$ are spacelike separated. Then we have:
\[
\mu(A \cap B \vert C) = \mu(A \vert C) \mu(B \vert C) \quad \forall C \in \Phi(\mathscr{P}_2(\mathscr{A},\mathscr{B}))
\]
Consider again the regions $\mathscr{X}$ and $\mathscr{Y}$ defined by eqs. \eqref{def:region_x}-\eqref{def:region_y}.

Now if $dom(A) \subset \mathscr{A}$ then we also have that $dom(A) \subset (\mathscr{A} \cup \mathscr{X})$. Similarly if $dom(B) \subset \mathscr{B}$ then $dom(B) \subset (\mathscr{B} \cup \mathscr{Y})$. This means that events $A$ and $B$ occur in regions $\{ (\mathscr{A} \cup \mathscr{X}), (\mathscr{B} \cup \mathscr{Y}) \}$ which, as before, can be seen to be space-like separated. Thus we can apply SO2 to these two regions. If we calculate the truncated joint past of the regions
$(\mathscr{A} \cup \mathscr{X})$ and $(\mathscr{B} \cup \mathscr{Y})$ we find however that it coincides with the mutual past of the regions $\mathscr{A}$ and $\mathscr{B}$, i.e.\
\begin{equation}\label{eq:crucial}
\mathscr{P}_2((\mathscr{A} \cup \mathscr{X}),(\mathscr{B} \cup \mathscr{Y}))=\mathscr{P}_1(\mathscr{A},\mathscr{B})
\end{equation}
This means that eq.\ \eqref{eq:so2} is the same as eq.\ \eqref{eq:so1}; hence SO2 $\Rightarrow$ SO1.
\end{description}

\section{Finite and infinite SO1 and SO2}
\label{sec:finite-infinite}
The equivalence of SO1 and SO2 does not seem to fit well with intuition: Existence of an explanation of a correlation in terms of a common cause system that is localized in the {\em mutual past} of the correlated events does not seem to be a priori inferable from an explanation in terms of common causes of which we only know that they are localized in the (truncated) {\em joint past}, which is typically a larger region; yet, the equivalence, if holds, entails that we are licensed to make such an inference. How can one reconcile the equivalence claim with this intuition?

The proof of the equivalence claim shows that the following two conditions are crucial in showing the equivalence:
\begin{description}
\item[Loose domain:] The screening-off conditions \eqref{eq:so1}-\eqref{eq:so2} in SO1 and SO2 are required {\em not} for the full specifications of the intersection and truncated joint past of the least domains $dom(A)$ and $dom(B)$ of the correlated events $A$ and $B$ but for the full specifications of the intersection and truncated past of {\em all} spacelike separated regions $\mathscr{A}$ and $\mathscr{B}$ containing the least domains $dom(A)$ and $dom(B)$, respectively.

\item[Tricky regions:] Because of the above {\bf Loose domain} property of the principles SO1 and SO2, in the proof SO2 $\Rightarrow$ SO1 it is allowed that SO2 is applied to regions of the form $(\mathscr{A} \cup \mathscr{X})$ and $(\mathscr{B} \cup \mathscr{Y})$ that have the peculiar property that their truncated joint past is equal to the mutual past of $\mathscr{A}$ and $\mathscr{B}$. This is achieved by taking $\mathscr{X}$ and $\mathscr{Y}$ as defined by eqs.\ \eqref{def:region_x}-\eqref{def:region_y}.
\end{description}

Henson himself notes \cite[p.\ 527]{Henson2005} that the {\bf Loose domain} property is somewhat unnatural and needs justification because the natural condition would be to require the screening-off conditions \eqref{eq:so1}-\eqref{eq:so2} to hold for the full specifications of the intersection and truncated joint past of the least domains of decidability $dom(A)$ and $dom(B)$ of the correlated events $A$ and $B$,  not for the full specifications of {\em all} regions $\mathscr{A}$ and $\mathscr{B}$ containing the least domains of decidability. The argument Henson provides for the justification is that if one requires the screening-off conditions for disjoint events $A_i$ all having least domain of decidability to be {\em equal} to $\mathscr{A}$ then this entails (by the properties of the least domain of decidability function $dom$) that the screening-off conditions hold for events of the form $A=\cup A_i$ and these events have their least domain of decidability just contained but not necessarily equal to $\mathscr{A}$. Henson points out that such a justification ``... can always be constructed for any $\mathscr{A}\supset dom(A)$ {\em as long as there are events with $\mathscr{A}$ as their least domain of decidability}''\cite[p.\ 527]{Henson2005} (our emphasis).

In particular such a justification is needed in case of the tricky domains $(\mathscr{A}\cup \mathscr{X})$ and $(\mathscr{B} \cup \mathscr{Y})$ that feature in the proof of the equivalence claims because these regions are strictly larger than the least domains of decidability $\mathscr{A}$ and $\mathscr{B}$. Note however that the regions $(\mathscr{A}\cup \mathscr{X})$ and $(\mathscr{B} \cup \mathscr{Y})$ are in a sense causally very large. For a typical example, take $\mathbb{S}$ to be the Minkowski spacetime $M$: Let $\mathscr{A}$ and $\mathscr{B}$ be two spacelike separated double cones. Then $J^-(\mathscr{A})\setminus \mathscr{A}$ is the infinitely large region of the full causal past of the double cone $\mathscr{A}$ minus
the finite double cone subtracted from it, and $(J^-(\mathscr{A})\setminus \mathscr{A})\setminus J^-(\mathscr{B})$ is the also infinitely large strip that remains of $J^-(\mathscr{A})\setminus \mathscr{A}$ after subtracting from it the full past of $\mathscr{B}$. It is then clear why the truncated joint past of $(\mathscr{A} \cup \mathscr{X})$ and $(\mathscr{B} \cup \mathscr{Y})$ is equal to the mutual past of the double cones $\mathscr{A}$ and $\mathscr{B}$ (see Fig.\ \ref{fig:jointpastAuXBuY}).

\begin{figure}
\begin{center}
\includegraphics{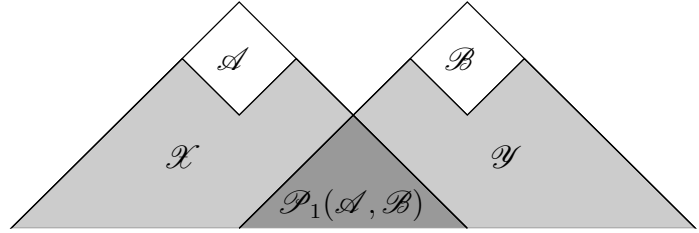}
\caption{The truncated joint past of $(\mathscr{A} \cup \mathscr{X})$ and $(\mathscr{B} \cup \mathscr{Y})$ is equal to the mutual past of $\mathscr{A}$ and $\mathscr{B}$}
\label{fig:jointpastAuXBuY}
\end{center}
\end{figure}

Adjectives ``finite'' and ``infinite'' in the previous sentences are used in the sense of a metric, which is available in Minkowski spacetime: a double cone $\mathscr{A}$ is finite in the sense of the metric, $(J^-(\mathscr{A})\setminus \mathscr{A})\setminus J^-(\mathscr{B})$ is not. But a metric is not assumed to be part of a general causal set, so finiteness and infiniteness cannot be defined in metric terms. On the basis of the example of double cones in Minkowski spacetime one can however formulate, in terms of the causal order only, what is important: subtracting a double cone from its causal past is nonempty; in addition, a double cone is causally closed in the sense of being equal to its causal closure, where causal closure of a spacetime region $O$ is, by definition, the set $(O')'$ where $O'$, the causal complement of $O$, is the set of points in the Minkowski spacetime that are spacelike from every point in $O$.
Furthermore, every bounded region $\mathscr{A}$ in Minkowski spacetime is contained in a double cone, the smallest of such double cones is equal to the causal closure of $\mathscr{A}$. The notion of causal complement and causal closure makes perfect sense in any causal set $\mathbb{S}$ in place of the Minkowski spacetime, and all this motivates the following definition.

\begin{defn}\label{def:causalfiniteness}
{\rm
A region $\mathscr{A}$ in $\mathbb{S}$ is called \emph{causally finite} if the causal past of its causal closure (denoted by $\bar{\mathscr{A}}$) is strictly larger than itself, i.e.\ if it holds that
\begin{equation}
J^-(\bar{\mathscr{A}})\setminus \bar{\mathscr{A}}\not=\emptyset
\end{equation}
The region is called \emph{causally infinite} if it is \emph{not} causally finite.
}
\end{defn}
As the case of Minkowski spacetime indicates the regions $(\mathscr{A}\cup \mathscr{X})$ and $(\mathscr{B} \cup \mathscr{Y})$ can be causally infinite in the sense of the above definition and since SO1 and SO2 are applied to them,  Henson's justification why this is conceptually admissible needs to be invoked, which, as we saw, assumes that there are events with $(\mathscr{A}\cup \mathscr{X})$ and $(\mathscr{B} \cup \mathscr{Y})$ as {\em least} domains of decidability.
Causally finite and causally infinite regions are conceptually very different however if considered as least domains of decidability: an event whose least domain of decidability is equal to a causally infinite set $\mathscr{Z}$ is only comprehendible by taking into account the complete causal structure of the region $\mathscr{Z}$. This whole region might very well be empirically inaccessible in its entirety however (think of the case of the double cones in Minkowski spacetime); hence one might not be in the position to decide empirically whether such an event has happened or not. In particular, observing a correlation involving such an event might very well be problematic.

Thus it is conceptually well-motivated to distinguish two versions of both SO1 and SO2: the original ones, in which no restrictions are imposed on the regions, call these principles {\em infinite} SO1 and {\em infinite} SO2, and ones in which  $\mathscr{A}$ and $\mathscr{B}$ are restricted to causally finite regions:

\begin{finiteso1}
For all events $A$ and $B$ with $dom(A) \subset \mathscr{A}$ and $dom(B) \subset \mathscr{B}$, if $\mathscr{A}$ and $ \mathscr{B}$ are causally finite space-like separated regions, then
\begin{align}
& \mu(A \cap B \vert C) = \mu(A \vert C) \mu(B \vert C) \quad \forall C \in \Phi(\mathscr{P}_1(\mathscr{A},\mathscr{B})).
\label{eq:finso1}
\end{align}
\end{finiteso1}
\begin{finiteso2}
For all events $A$ and $B$ with $dom(A) \subset \mathscr{A}$ and $dom(B) \subset \mathscr{B}$, if $\mathscr{A}$ and $\mathscr{B}$ are causally finite space-like separated regions then:
\begin{align}
& \mu(A \cap B \vert C) = \mu(A \vert C) \mu(B \vert C) \quad \forall C \in \Phi(\mathscr{P}_2(\mathscr{A},\mathscr{B})).
\label{eq:finso2}
\end{align}
\end{finiteso2}

\section{Are the four causality principles equivalent?\label{sec:equivalent?}}
What is the relation of the four causality principles {\em finite} and {\em infinite} SO1 and {\em finite} and {\em infinite} SO2?

As we recalled in Section~\ref{sec:CCP}, Henson has shown that {\em infinite} SO2 entails {\em infinite} SO1. As the example of double cones in Minkowski space shows, the region $(\mathscr{A} \cup \mathscr{X})$,
where $\mathscr{X}$ is defined by eq.\ \eqref{def:region_x}, may not be causally finite even if both $\mathscr{A}$ and $\mathscr{B}$ are. Consequently, the proof presented in Section~\ref{sec:CCP} of the implication {\em infinite} SO2 $\Rightarrow$ {\em infinite} SO1, which relies in a crucial way on applying SO2 to regions of the form $(\mathscr{A} \cup \mathscr{X})$, does {\em not} prove the implication {\em finite} SO2 $\Rightarrow$ {\em infinite} SO1, not even the implication {\em finite} SO2 $\Rightarrow$ {\em finite} SO1.

What Henson's argument in Section~\ref{sec:CCP} in favor of the implication SO1 $\Rightarrow$ SO2 shows, strictly speaking, is that, given any pair $A,B$, and $C\in\Phi (\mathscr{P}_1(\mathscr{A},\mathscr{B}))$ satisfying \eqref{eq:so1}, there is always a $C'\in\mathscr{P}_2(\mathscr{A},\mathscr{B})$ (namely $C' = C \cap X \cap Y$) such that $A$, $B$ and $C'$ satisfy the condition required by SO2. This does not yield yet that SO1 entails SO2, however: in order to infer this, one needs to show that these $C' \in \mathscr{P}_2(\mathscr{A},\mathscr{B})$ events exhaust the set $\Phi (\mathscr{P}_2(\mathscr{A},\mathscr{B}))$ of {\em all\/} possible full specifications of $\mathscr{P}_2(\mathscr{A},\mathscr{B})$ -- and it remains unclear why this is the case, especially in case of causally infinite least domains of decidability. Thus we regard the problem of whether the implication [(finite or infinite) SO1 $\Rightarrow$ (finite or infinite) SO2] as open but likely true and intuitively plausible.

The implications [infinite SO1 $\Rightarrow$ finite SO1] and
[infinite SO2 $\Rightarrow$ finite SO2] are trivial. The diagram in Fig.\ \ref{fig:logical_diagram} summarizes the logical relations known between these causality principles at this time, question marks indicating open problems.
\begin{figure}
\begin{center}
\includegraphics{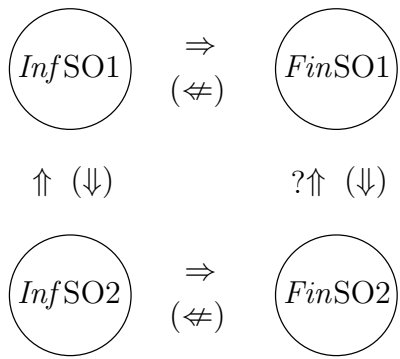}
\caption{Logical relations between the diverse screening-off principles. Relations in parenthesis are considered as likely to hold, although no complete formal proof has been found. A question mark indicates that the corresponding logical relation is not known.}
\label{fig:logical_diagram}
\end{center}
\end{figure}

The truth of the claim formulated above that Henson's proof in Section~\ref{sec:CCP} of the implication [{\em infinite} SO2 $\Rightarrow$ {\em infinite} SO1] does {\em not} prove either of the implications [{\em finite} SO2 $\Rightarrow$ {\em infinite} SO1] or [{\em finite} SO2 $\Rightarrow$ {\em finite} SO1] does {\em not} entail that none of these implications is true, and we do not have a formal proof that these implications do not hold. But the analysis in local algebraic relativistic quantum field theory (AQFT) of the status of SO1 and SO2 yields a strong argument that \textit{finite} SO2 is strictly weaker than \textit{infinite} SO2. So we conclude by commenting on the status of the causality principles SO1 and SO2 in local algebraic relativistic quantum field theory (AQFT).

It was shown in \cite{Redei-Summers2002}, \cite{Redei-Summers2005corr} (see also \cite{Redei1997}, chapter in \cite{Redei1998} and \cite{Redei2002Krakko}) that if AQFT satisfies the condition of local primitive causality, then {\em finite} SO2 holds in AQFT in the specific sense that there is a single common cause of spacelike correlations predicted by faithful states between projections contained in algebras pertaining to spacelike separated double cones and the common cause is localized in the truncated joint past of the backward light cones of the double cones containing the correlated projections (Proposition~3 in \cite{Redei-Summers2002}). This fact was referred to by saying that AQFT satisfies the {\em weak common cause principle}. It could not be proved so far however that AQFT also satisfies (even the {\em finite}) SO1. In view of the distinctions made here between finite and infinite versions of SO1 and SO2, it is instructive to see why the proof in \cite{Redei-Summers2002} of the claim that AQFT satisfies the weak common cause principle is too weak to prove that {\em infinite} SO2 holds in AQFT, because the proof shows how radically different finite and infinite versions of SO2 and SO1 can be.

The proof of {\em finite} SO2 in AQFT consists of the following steps:
\begin{description}
\item[step 1]
One shows, utilizing the special {\em type} of the local algebras pertaining to double cones that the probability theories describing the observables represented by these algebras have the particular measure theoretic feature of being atomless.
\item[step 2] One proves that atomlessness entails that there exists a projection in the smallest double cone $\mathscr{C}$ containing the double cones $\mathscr{A}$ and $\mathscr{B}$ which satisfies the probabilistic conditions required of a common cause.
\item[step 3]
Using the local primitive causality condition one ``pushes back'' the common cause in $\mathscr{C}$ far enough back in the joint causal past of $\mathscr{A}$ and $\mathscr{B}$ until it is in the {\em truncated} joint past of $\mathscr{A}$ and $\mathscr{B}$.
\end{description}
(Recall that the local primitive causality condition states that the local algebra ${\cal A}(V)$ of observables associated with an open bounded spacetime region $V$ is equal to the local algebra of observables ${\cal A}(\bar{V})$ where $\bar{V}$ is the causal closure of $V$.)

This proof cannot be extended to conclude that {\em infinite} SO2 also holds in AQFT, for the following reasons:
\begin{itemize}
\item While it could be proved that local algebras associated to double cone regions have the measure theoretic atomlessness as a consequence of such algebras being type III von Neumann algebras, this latter algebraic feature is only known to hold in AQFT for algebras pertaining to special spacetime regions and hence cannot be assumed for arbitrary regions such as \eqref{def:region_x}-\eqref{def:region_y}. That is to say, already {\bf step 1} is not known to be valid for arbitrary, in particular causally infinite, regions.\\
    Moral: the type of measure space describing events may very well restrict the possible shapes of regions that are least domains of decidability for the events in the probability space.
\item {\bf Step 3} only works because double cones are causally finite (in the sense of Definition \ref{def:causalfiniteness}): Even if we could make {\bf step~1} for regions $\mathscr{A}\cup \mathscr{X}$  and $\mathscr{B}\cup \mathscr{Y}$ and did find a common cause in the {\em untruncated} joint past of $\mathscr{A}\cup \mathscr{X}$ and  $\mathscr{B}\cup \mathscr{Y}$ ($\mathscr{X}$ and $\mathscr{Y}$ defined by \eqref{def:region_x} and \eqref{def:region_y}), we would not be able to ``push back'' this common cause far enough in the untruncated joint past of $\mathscr{A}\cup \mathscr{X}$ so that it gets in the {\em truncated} joint past of $\mathscr{A}\cup \mathscr{X}$ and $\mathscr{B}\cup \mathscr{Y}$ -- precisely because these regions are causally infinite and thus there is no open bounded (hence causally finite) region $V$ having the property that its causal closure $\bar{V}$ is equal to the untruncated joint past of $\mathscr{A}\cup \mathscr{X}$ and $\mathscr{B}\cup \mathscr{Y}$.
\end{itemize}
Thus the situation in AQFT indicates that {\em finite} and {\em infinite} SO2 are {\em not} equivalent and thus that the logical relationships among these different causality principles is a very subtle and complicated matter, clarification of their relation requiring further work.

\section*{Acknowledgements}
Research supported by the Hungarian Scientific Research Fund (OTKA), contract number: K100715; and by the Spanish Ministry of Science and Innovation (MICINN) Research Project FFI2008-06418-C01-03.

\renewcommand{\refname}{References}
\pagestyle{plain}
\bibliographystyle{amsalpha}
\bibliography{RedeiBib}

\providecommand{\bysame}{\leavevmode\hbox to3em{\hrulefill}\thinspace}
\providecommand{\MR}{\relax\ifhmode\unskip\space\fi MR }
\providecommand{\MRhref}[2]{%
  \href{http://www.ams.org/mathscinet-getitem?mr=#1}{#2}
}
\providecommand{\href}[2]{#2}
\begin{thebibliography}{HSV12}

\bibitem[Hen05]{Henson2005}
J.~Henson, \emph{Comparing causality principles}, Studies in the History and
  Philosophy of Modern Physics \textbf{36} (2005), 519--543.

\bibitem[HSR04]{Hofer-Redei2004}
G.~Hofer-Szab\'o and M.~R\'edei, \emph{Reichenbachian common cause systems},
  International Journal of Theoretical Physics \textbf{43} (2004), 1819--1826.

\bibitem[HSR06]{Hofer-Redei2006}
\bysame, \emph{Reichenbachian common cause systems of arbitrary finite size
  exist}, Foundations of Physics Letters \textbf{35} (2006), 745--746.

\bibitem[HSV12]{Hofer-Vecsernyes2012}
G.~Hofer-Szab\'o and P.~Vecserny\'es, \emph{Reichenbach's common cause
  principle in algebraic quantum field theory with locally finite degrees of
  freedom}, Foundations of Physics \textbf{42} (2012), 241--255.

\bibitem[R{\'e}d97]{Redei1997}
M.~R{\'e}dei, \emph{Reichenbach's {C}ommon {C}ause {P}rinciple and quantum
  field theory}, Foundations of Physics \textbf{27} (1997), 1309--1321.

\bibitem[R{\'e}d98]{Redei1998}
\bysame, \emph{Quantum logic in algebraic approach}, Fundamental Theories of
  Physics, vol.~91, Kluwer Academic Publisher, 1998.

\bibitem[R{\'e}d02]{Redei2002Krakko}
\bysame, \emph{Reichenbach's common cause principle and quantum correlations},
  Modality, Probability and Bell's Theorems (T.~Placek and J.~Butterfield,
  eds.), NATO Science Series, II., vol.~64, Kluwer Academic Publishers,
  Dordrecht, Boston, London, 2002, pp.~259--270.

\bibitem[Rei56]{Reichenbach1956}
H.~Reichenbach, \emph{The direction of time}, University of California Press,
  Los Angeles, 1956.

\bibitem[RS02]{Redei-Summers2002}
M.~R\'edei and S.J. Summers, \emph{Local primitive causality and the common
  cause principle in quantum field theory}, Foundations of Physics \textbf{32}
  (2002), 335--355.

\bibitem[RS07]{Redei-Summers2005corr}
\bysame, \emph{Remarks on causality in relativistic quantum field theory},
  International Journal of Theoretical Physics \textbf{46} (2007), 2053--2062.

\bibitem[Sor03]{Sorkin2003}
R.D. Sorkin, \emph{Causal sets: Discrete gravity ({N}otes for the {V}aldivia
  {S}ummer {S}chool)}, 2003, \url{http://arxiv.org/abs/gr-qc/0309009}.

\bibitem[Uff99]{Uffink1999}
J.~Uffink, \emph{The principle of the common cause faces the {B}ernstein
  paradox}, Philosophy of Science, Supplement \textbf{66} (1999), 512--525.

\bibitem[WM10]{Wronski-Marczyk2010}
L.~Wronski and M.~Marczyk, \emph{Only countable {R}eichenbachian common cause
  systems exist}, Foundations of Physics \textbf{40} (2010), 1155--1160.

\end{thebibliography}

\end{document}